\documentclass[a4paper,twocolumn]{mwart}
\usepackage{graphicx}
\usepackage{amsmath}

\begin{document}
\title{Efficient method of designing optically-pumped vertical external cavity surface emitting lasers having equally excited
quantum wells}
\author{Micha\l{} Wasiak\thanks{Photonics Group, Institute of Physics, Technical University of \L\'od\'z, ul.~W\'olcza\'nska 219, 90-924 \L\'od\'z, Poland. e-mail: \texttt{michal.wasiak@p.lodz.pl}}}
\maketitle
\begin{abstract}
Even distribution of  carriers allows to maximize
optical gain of the Optically-Pumped Vertical External Cavity 
Surface Emitting Laser. In this paper we show how to distribute
the quantum wells and blocking layers in order to compensate the
exponential decay of the pumping beam intensity. Our model says whether
it is possible at all (for an assumed length of the device) and,
if it is, allows to find  positions of the blocking layers.
No iterations nor numerical calculations more sophisticated than
a standard calculator can do are required to use the model.
\end{abstract}
\section{Introduction}
The Optically Pumped Vertical External Cavity Surface Emitting Lasers
(OP-VECSELs) are able to emit high quality beams of multi-watt 
powers~\cite{Rudin:08}, thus combining the most important advantages
of Edge-Emitting and Vertical Cavity Surface Emitting Lasers 
(EELs and VCSELs). Moreover their external cavity may contain
additional elements like non-linear crystals or semiconductor saturable 
absorber mirrors (SESAMs), which
may be used for instance for frequency doubling and short pulse
generation. Usually, optical gain is provided by several quantum-well
(possibly double- or multi-quantum-well)
active regions, located at the successive anti-nodes of the standing wave.
The cavity is formed by a Distributed Bragg Reflector (DBR)
and an external mirror~\cite{Kuzn}.

The pumping beam penetrates the device in the direction 
nearly perpendicular to the layers. Absorption of the pumping beam
generates  carriers necessary to achieve  optical gain,
but also causes exponential decay of the intensity in the deeper
regions. If we simply placed, at each anti-node, the same number of
wells, the gains provided by the active regions would differ 
significantly. Since optical gain is a concave function
of carrier concentration (roughly $\sim\log(\mathcal{N}/\mathcal{N}_0)$, 
where $\mathcal{N}$ is the carrier concentration and $\mathcal{N}_0$
is the transparency concentration), the highest possible total gain
(for a fixed, arbitrary number of the carriers) is highest when all
the wells provide the same material gain. If the temperatures
of the wells can be assumed to be equal, we should
try to have the same carrier concentration in each well. The temperature
rise in the VECSEL can be high, most of the
temperature drop takes place in the substrate and DBR as these
part are much thicker than the active part. Generally, we want
to direct the same number of carriers to all the wells.

In order to do so, we have to increase the volume from which the distant
wells collect the carriers or increase the number of the wells in
the stronger pumped regions. To do the first thing one can use
so called \textit{blocking layers}---thin wide-gap layers which
block  carrier diffusion (in case of AlAs blocking layers in GaAs, 
the thickness of a few nanometers is sufficient to block the carrier 
diffusion). They define the volume from which
the well(s) between them collect the carriers.~\cite{Moroz}.
The question how to place them is not trivial, mainly because
positions of the wells are restricted to the anti-nodes, which
strongly restricts  positions of the blocking layers. 

If there are two or more anti-nodes between two subsequent blocking 
layers, one has to take into account the carrier diffusion in order
to find the actual carrier concentration in the two (or more)
active regions. This makes the analysis more complicated~\cite{Moroz}.
Our goal is to build an analytical model which allows to construct the desired
scheme without using complicated calculation.

\section{The model}
\begin{figure*}[t]
  \centering
  \includegraphics{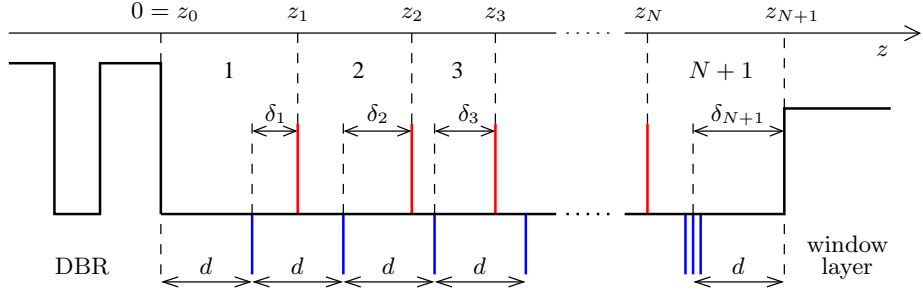}
  \caption{Scheme of wells and blocking layers distribution in a VECSEL. Wells are blue (downwards), blocking layers red (upwards). Numbers $1,2,\dots,N+1$ denote the segments.}
  \label{fig:sch}
\end{figure*}

As we mentioned, in order to avoid consideration of carrier
diffusion, we restrict our interest to the case, where in each segment 
(area bounded by the neighbouring blocking layers) there is only one 
active region (we treat the DBR as the $0$th blocking layer, see 
Fig.~\ref{fig:sch}), and the active regions are placed at each anti-node.
The window layer acts as the last blocking layer and due to the optical
reasons must be located at an anti-node. This means that the last
segment must be thicker than $d$ (see Fig.~\ref{fig:sch}). Therefore, 
in order to have the same number of generated carrier per a QW, we have to
put more wells in this segment and treat it in a different manner in our calculations.

In order to obtain a handy result we base our model on the following 
simplifying assumptions:
\begin{enumerate}
\item Widths of the wells and blocking layers are negligible. More
precisely, their presence (the blocking layers do not absorb
the pump, on the other hand the wells have higher absorption than the adjacent
bulk material) introduces a tolerable error. 
\item The blocking layers block totally carrier diffusion
\item We neglect reflection of the pumping light from the DBR
  and from the blocking layers
\item Carrier losses in the absorbing barrier are negligible
  (relative to the losses in the QWs)
\end{enumerate}

In Fig.~\ref{fig:sch} a scheme of the VECSEL is presented. Distance $d$
must be a multiple of a half of the emitted wavelength. Usually $d\approx\lambda/(2n_r)$,
where $\lambda$ is vacuum wavelength and $n_r$ is refractive index of the 
barrier. Small deviations from the exact equality come from the presence of the 
wells and blocking layers. 
In our scheme the pumping light comes from the right, so
under our assumptions the pumping wave intensity is 
described by the following formula:
\begin{equation}
  \label{eq:pump}
  I(z) = I_0\exp(\alpha z)
\end{equation}
where $\alpha$ is the absorption in the barriers, $I_0$ is a normalisation
constant, defining the pump power. Number of the carriers generated
in $n$-th segment is simply
\begin{equation}
  \label{eq:pown}
  P_n=I_0 \big(\exp(\alpha z_n) - \exp(\alpha z_{n-1})\big)
\end{equation}
We assume that in segments $1,2,\dots,N$ there are the same number of wells 
(in this paper---just one well), and in the last segment, $N+1$,
we put $k$ wells. 
we want to distribute the 
blocking layers such that 
\begin{equation}
  \label{eq:pow}
  P_1=P_2=\dots = P_N = \frac{1}{k}P_{N+1}
\end{equation}
As we have $k$ times more wells in the last segment, we want to have $k$
times more carriers generated there. Possible values of numbers $N$ and $k$
will be determined in our calculations.
Position $z_n$ can be written as (see Fig.~\ref{fig:sch})
\begin{equation}
  \label{eq:z}
  z_n = nd + \delta_n
\end{equation}
Then we have
\begin{align*}
  P_1 &= I_0\big[\exp\big(\alpha(d +\delta_1)\big)-1\big] =\\
  &=I_0\big[\exp(\alpha d)\exp(\alpha\delta_1)-1\big]\\
  P_2 &= I_0\big[\exp\big(\alpha(2d +\delta_2)\big)-[\exp\big(\alpha(d +\delta_1)\big)\big] =\\
  &=I_0\big[\exp(2\alpha d)\exp(\alpha\delta_2)-\exp(\alpha d)\exp(\alpha\delta_1)\big]\\
  P_3 &= I_0\big[\exp\big(\alpha(3d +\delta_3)\big)-[\exp\big(\alpha(2d +\delta_2)\big)\big] =\\
  &=I_0\big[\exp(3\alpha d)\exp(\alpha\delta_3)-\exp(2\alpha d)\exp(\alpha\delta_2)\big]
\end{align*}
Let us introduce the following symbols:
\begin{equation}
  \label{eq:symb}
  a=\exp(-\alpha d)\qquad x_n = \exp(\alpha\delta_n)
\end{equation}
Note that $0<a<1$, regardless of $\alpha$ and $d$.
Using these symbols we can write the system of equations \eqref{eq:pow} as:
\begin{align}
  \label{eq:system_caly}
  \notag
  0 &= x_2 - 2ax_1 + a^2  \\ \notag
  0 &=x_3 - 2ax_2 + a^2x_1 \\ \notag
  0 &=x_4 - 2ax_3 + a^2x_2 \\ 
  & \vdots \\ \notag
  0 &=x_{N} -2ax_{N-1} + a^2x_{N-2} \\ \notag
  0 &=x_{N+1} -(k+1)ax_{N} + ka^2x_{N-1} \\ \notag
  \frac{1}{a} &=x_{N+1} \qquad\text{because }\delta_{N+1} = d \notag
\end{align}
As one can see, thanks to the extraordinary properties of 
exponential function we got something as simple as a system of
$N+1$ linear equations with $N+1$ unknowns. Because 
we do not know what are the values of $N$ and $k$, it is convenient
to consider first only equations concerning first $N$ segments, 
i.e.~those defined by the blocking layers which position we can
choose. Thus we consider the system of $N-1$ equation with $N$ unknowns:
\begin{align}
  \label{eq:system}
  \notag
  x_2 - 2ax_1 + a^2 &= 0 \\ \notag
  x_3 - 2ax_2 + a^2x_1 &= 0 \\ \notag
  x_4 - 2ax_3 + a^2x_2 &= 0 \\ 
  & \vdots \\ \notag
  x_{N} -2ax_{N-1} + a^2x_{N-2} &= 0 \\ \notag
\end{align}
If we treat one of the unknowns as a parameter we can solve the 
system. As the parameter we choose $x_1$ as the first blocking layer
must be always present. This way we obtain:
\begin{align}
  \label{eq:sol}
  \notag
  x_2 &= a(2x_1-a)\\ \notag
  x_3 &= a^2(3x_1-2a)\\ 
  &\vdots \\ \notag
  x_N &= a^{N-1}\big(Nx_1-(N-1)a\big)\\ \notag
\end{align}
Although the system~\eqref{eq:system} has always the solution
in real numbers, we must check if the solution fulfils 
the additional conditions, i.e.:
\begin{equation}
  \label{eq:condn}
  1 < x_n < 1/a \quad \forall n = 1,2,\dots,N
\end{equation}
The above conditions say simply that $0 < \delta_n < d$. It assures
that in each segment there is one active region. 

Substituting \eqref{eq:sol} to \eqref{eq:condn} we get:
\begin{align}
  \label{eq:ineq}
  \notag
  1 &< x_1 < \frac{1}{a} \\ \notag
  \frac{1}{2}\left(a+\frac{1}{a}\right) &< x_1 < \frac{1}{2}\left(a+\frac{1}{a^2}\right) \\ \notag
  \frac{1}{3}\left(2a+\frac{1}{a^2}\right) &< x_1 < \frac{1}{3}\left(2a+\frac{1}{a^3}\right) \\
  &\vdots \\ \notag
  \frac{1}{N}\left((N-1)a+\frac{1}{a^{N-1}}\right) &< x_1 < \frac{1}{N}\left((N-1)a+\frac{1}{a^{N}}\right) \\ \notag
\end{align}
If the above inequalities are inconsistent, it is impossible to build 
a system of $N$ active regions (of the same number of QWs), having
equal carrier concentrations.

The above system can be significantly simplified. Let us denote
\begin{gather}
  \label{eq:left}
  L_n=\frac{1}{n}\left((n-1)a + \frac{1}{a^{n-1}}\right)\\
  R_n = \frac{1}{n}\left((n-1)a+\frac{1}{a^n}\right)
\end{gather}
being just the left and right hand side of the $n$-th inequality. 
Basic calculations show that:
\begin{equation}
  \label{eq:ldif}
  L_{n+1} - L_{n} = \frac{1}{n(n+1)a^n}\left(a^{n+1} - (n+1)a + n\right) 
\end{equation}
The sign of the difference is determined by polynomial $l_n(a) = a^{n+1} - (n+1)a + n$.
It is easy to see that $l_n(1) = 0$, $l_n'(a) = (n+1)(a^n-1) \leq 0 \ \forall a\in[0,1]$.
It means that $l_n(a) \leq 0\ \forall a\in[0,1]$ (in our case $0<a<1$), and 
hence
\begin{equation}
  \label{eq:leftN}
  L_1<L_2<\dots<L_N
\end{equation}
Thanks to that property we can replace all the inequalities $L_n<x_1, n=1,2,\dots,N$ 
by one: $L_N<x_1$. There is no similar properties of the right hand sides.
Now we can write the system of inequalities~\eqref{eq:ineq} in a more compact
way:
\begin{multline}
  \label{eq:ineqc}
  \frac{1}{N}\left((N-1)a+\frac{1}{a^{N-1}}\right) < x_1 < \\ \min_{n=1,\dots,N}\left(\frac{1}{n}\left((n-1)a+\frac{1}{a^n}\right)\right) = \mathcal R(N)
\end{multline}
Which of the right hand sides is the actual minimum depends on $a$ 
(i.e.~on the material absorption and distance between the active regions).

If for a certain $N$ $L_N\geq \mathcal R(N)$, it is impossible to
build $N$ (and any greater number, since $L_N$ is increasing and 
$\mathcal R(N)$ is non-increasing with $N$) equally pumped segments. 
This way we can
find the possible values of $N$. When we choose one of them, we can
return to the full system of equations~\eqref{eq:system_caly}. Solving
it we can find values of all $x_1,x_2,\dots,x_N$, but now we are interested 
only in $x_1$:
\begin{equation}
  \label{eq:iks1}
  x_1 = \frac{1}{(N+k)a^{N+1}} + a\frac{N+k-1}{N+k}
\end{equation}
If, for the chosen $N$, we can find a natural number $k$ such that $x_1$ 
calculated with the above formula fulfils condition:
\begin{equation}
  L_N < x_1 < \mathcal R(N)
\end{equation}
we know that we can build $N$ one-well segments concluded by one $k$-well
segment, having equally pumped quantum wells. If there is no such $k$, or
it is too high from the practical point of view, one has to decrease number
$N$ and repeat the procedure.

Finally, when we have suitable $N$, $k$ and hence $x_1$, we can
calculate all the other $x_2,x_3,\dots,x_N$ using~\eqref{eq:sol} and
then the actual positions $z_n, n=1,2,\dots,N+1$ of the blocking layers
using formulae~\eqref{eq:z} and~\eqref{eq:symb}.
\section{Example}\label{sec:E}
Let us consider an important example: a VECSEL with GaAs barirers,
emitting at $980\,$nm, pumped by a $808\,$nm laser. Assuming 
$\alpha = 1.3\cdot 10^4\,1/\mathrm{cm}$~\cite{absorp}, and $n_r=3.52$ we get the following parameters:
\begin{equation}
  d \approx 139.2\,\mathrm{nm}\qquad a\approx 0.8345
\end{equation}
Of course in a real calculations one should use more precise values,
because the optical properties of the structure are more sensitive
to the distances between the wells.
First we will check whether it is possible to have seven one-well
segments. We calculate, using~\eqref{eq:left}:
\begin{equation}
  L_7 = 1.13837 \qquad \mathcal R(7) = \min_{n=1,\dots,7} R_n = R_3 = 1.12997
\end{equation}
We see that $L_7 > \mathcal R(7)$, so we cannot construct seven such segments.
But since
\begin{equation}
  L_6 = 1.10730 < \mathcal R(6) = R_3 = 1.12997
\end{equation}
we can find out if we can close such 6-segment sequence by the multi-well
one. Using~\eqref{eq:iks1} for $N=6$ we calculate:
\begin{align}
  x_1(k=3) = 1.13612 &> \mathcal R(6)\\
  x_1(k=4) = 1.10595 &< L_6
\end{align}
Since $x_1$ is a decreasing function of $k$, there is no possibility
to find a suitable $k$. It means that we cannot build the final
segment with an integer number of wells such that it has the
same carrier concentration as in all the others. The non-integer
values of $k$ could have physical meaning as $k$ is actually the
ratio between number of wells in the last segment and number of wells 
in the other ones. If we assumed the other segments to contain not one, but 
two wells each, we could consider $k=3.5$. But it would mean that we
have to put as many as seven quantum wells in the last segment, which
is too many, because the peripheral wells would be placed far from the 
anti-node of the standing wave.

If not $6$, let us try $N=5$. Now we have 
$L_5 = 1.08005$, $\mathcal R(5) = 1.12997$. For the new~$N$:
\begin{align}
  x_1(k=2) = 1.13837 &> \mathcal R(5)\\
  L_5 < x_1(k=3) = 1.10038 &< \mathcal R(5)\\
  x_1(k=4) = 1.07083 &< L_5
\end{align}
Number $x_1(k=3)$ fulfils all the conditions, so we can build
$5$ one-well segments and one $3$-well segment on the top, with
all the $8$ wells equally pumped. 
Now we calculate
$x_2,\dots,x_N$ using~\eqref{eq:sol}, extract $\delta_n$ given by:
\begin{equation}
  \label{eq:deltan}
  \delta_n=\frac{\log(x_n)}{\alpha}
\end{equation}
Finally we get $z_1,\dots,z_{N+1}$ from formula~\eqref{eq:z}.
Positions of the wells 
(which are independent on  $x_1$) and blocking layers are presented in 
table~\ref{tab:pos}. One should remember that positions of the wells ($nd$)
and hence numbers $z_n$ are only approximations. The reliable values
are $\delta_n$---positions of the blocking layers relative to the adjacent
quantum well (see Fig.~\ref{fig:sch}).
 \begin{table}[h]
   \caption{Positions of the wells and blocking layers in the 6-segment scheme,
     rounded to full nanometers}
   \label{tab:pos}
   \centering
   \renewcommand{\arraystretch}{1.2}
   \medskip
   \begin{tabular}[tab:pos]{|c|l|r|r|r|r|}
     \cline{3-6}
     \multicolumn{2}{c}{}&\multicolumn{4}{|c|}{[nm]}\\
     \hline
     $n$ &\multicolumn{1}{c|}{$x_n$} & $\delta_n$ & $nd$ & $z_n$ & $(n+1)d$\\
     \hline
     1 & 1.10038 & 74 & 139 & 213 & 278\\
     2 & 1.14012 & 101 & 278 & 379 & 418\\
     3 & 1.13656 & 98 & 418 & 516 & 557\\
     4 & 1.10293 & 75 & 557 & 632 & 696\\
     5 & 1.10578 & 37 & 696 & 733 & 835\\
     6 & 1.19837 & 139 & 835 & 974 & 974\\
     \hline
   \end{tabular}
 \end{table}

Looking at the values in the table one can see that the lowest
distance between a well and a blocking layer is $37$\,nm. This
is a safe distance from the technological point of view,
even taking into account non-zero thicknesses of the wells and
the blocking layers. 
The total thickness of the
absorbing area is $7d$, $\exp(-7d\alpha)\approx 0.28$, and hence 
over $70\%$ of the pumping power is absorbed. As  the blocking
layers are generally located near the nodes of the standing wave,
and  their thickness can be as low as a few nanometers, 
their presence modifies the optical properties of the resonator 
in a very limited degree.
\section{Summary}
We have shown an efficient and simple way to design
the VECSEL structure such that the carrier concentration,
and hence  optical gain (with the assumption that the 
temperature differences between the wells are not high),
is equal in all the quantum wells. This configuration
gives the highest modal gain for given number of carriers,
so is highly desirable.

We presented an example of a GaAs-based structures with
6 active regions in subsequent anti-nodes of the standing
wave. This is the longest possible design in which one use
1-well segments except the last one. Our calculations can
be easily modified to describe segments with different number 
of wells. In this case longer absorbing areas can be achieved,
if necessary. However, the longer the area is, the higher
temperature differences appear there, which spoils the desired 
gain uniformity.
\section{Acknowledgement}
This work was partially supported by COST Action MP0805.
\bibliographystyle{elsarticle-num}
\bibliography{VECSEL_arXiv}
\end{document}